\documentstyle [12pt]{article}
\begin{document}
%
%
\newcommand{\Abs}[1]{|#1|}
\newcommand{\EqRef}[1]{(\ref{eqn:#1})}
\newcommand{\FigRef}[1]{fig.~\ref{fig:#1}}
\newcommand{\Abstract}[1]{\small
   \begin{quote}
      \noindent
      {\bf Abstract - }{#1}
   \end{quote}
    }
\newcommand{\FigCap}[2]{
\ \\
   \noindent
   Figure~#1:~#2
\\
   }
%
%
%
%
\title{Resonance spectra and a periodic orbit sum rule for bound chaotic
systems }
\author{Per Dahlqvist \\
Mechanics Department \\
Royal Institue of Technology, S-100 44 Stockholm, Sweden\\[0.5cm]
}
\date{}
\maketitle

\ \\ \ \\

%
\Abstract{We consider the spectrum of the evolution operator for bound
chaotic systems by evaluating its trace. This trace
is known to approach unity as $t \rightarrow \infty$
for bound systems. It is written as the
Fourier transform of the logaritmic derivative of a
zeta function whose zeros are identified with the
eigenvalues of the operator.
Zeta functions are expected to be entire only for very specific systems,
like  Axiom-A systems. For bound chaotic systems complications arise due
to e.g. intermittency and non completeness of the symbolic dynamics.
For bound intermittent system an approximation of the zeta function is
derived. It is argued that bound systems
with {\em long time tails} have branch cuts in the zeta function and traces
approaching unity as a powerlaw. Another feature of bound chaotic systems,
relevant for the asymptotics of the trace, is that the dominant time scale
can be much longer than the period of the shortest periodic orbit.
This the case for e.g. the hyperbola billiard.
Isolated zeros of the zeta function for the
hyperbola billiard are  evaluated by means of a cycle expansion.
Crucial for the success of this approach is the identification of a sequence
of periodic orbit responsible for a logaritmic
branch cut in the zeta function.
Semiclassical implications are discussed at the
end.}

\newpage

\section{Introduction}

Time correlations provide an efficient tool for the analysis of dynamical
systems. For, some,
chaotic systems
correlations decay exponentially, reflecting loss of memory during
evolution.
Often the correlation function is found to be modulated
by some typical frequence(s) related to the periodic orbits of the system.
Its Fourier transform will thus have
poles ({\em resonances})
in the complex
frequence ($k$) plane. The imaginary parts correspond to the decay rates
and the real parts to the oscillation frequences.
Integrable systems have resonances {\em on} the real axis, reflecting
quasiperiodicity of the motion.

A rigorous theory for resonances
has so far been developed only for Axiom-A systems
\cite{Poli,Rue1,Rue2,Rugh}.
The correlation function is then
meromorphic (i.e. it has nothing worse than poles) in a strip around the real
$k-$axis. For mixing Axiom-A maps one can establish exponential bounds on the
decay of correlations, i.e. there is a strip (gap) above the real axis
without resonances. But even for Axiom-A flows the resonances may lie
arbitrarily close to the real axis leading to non-exponential decay of
correlations \cite{Viv1,Viv2}.
It is a general experience that the positions of
the resonances are independent
of the particular observables whereas the residues are not.
This is proven only for Axiom-A systems but numerical work suggest that it
holds more generally.
This is very intriguing since one can thus characterize a
system by its (classical) resonance spectrum,
in nice analogy with its quantum spectrum.

However, Axiom A systems are very idealistic systems from a physical point of
view, although they may be realized by e.g. some open scattering systems.
The purpose of the present paper is to
study resonance spectra in {\em bound}
Hamiltonian systems, exhibiting chaos.
Then we know that there is a trivial pole at $k=0$ reflecting the boundedness
of the system. All other resonances lie in the upper halfplane. This is
essentially all we know for certain.

Crucial for the succesful theory of Axiom-A system is the existence of a
finite markov partition.
One can thus perform a well converging cycle expansion of a zeta function
in order to locate
the resonances \cite{Rugh}.
Such finite grammar does generally not exist
for bound systems. This leaves us without any a priori knowledge of the
analytical structure of the zeta function we are going to study.

We will restrict ourselves to ergodic billiards in the Euclidean plane.
Seemingly all
systems of this type have one thing in common; they are intermittent.
Typically,
the motion will intermittently alternate between a chaotic region
of phase space and a regular one.  This
might  imply power law decay of correlations.
In a paper by Baladi, Eckmann and Ruelle \cite{BER}
(hereafter referred to as the
BER approximation) it is demonstrated how the approximate positions of the
resonances for intermittent systems may be calculated. This result is directly
applicable to a wide class of bound chaotic systems and we will make use of
this observation in the following.

The second main theme of the paper is a periodic orbit sum rule discovered
in \cite{Ozo}. It is thus shown that a certain
weighted sum over all periodic orbits
up to period $t$ tends to unity as $t \rightarrow \infty$.
This is an asymptotic result and it is of course highly interesting to know
how this asymptotic limit is approached.
The sum rule is a crucial ingredient when computing
correlations in quantum spectra.

The paper is outlined as follows. We begin, in section 2.1, by
reviewing the derivation of the classical zeta function whose zeros are
associated with the resonances.
Then, in section 2.2, we review how, according to the BER approximation,
the positions of the resonances are related to a simple probability
distribution associated with the system. In section 3 we relate, by
heuristic arguments, different features of this probability
distribution to the resonance spectrum. In section 4 we apply cycle
expansions of the zeta function to a specific dynamical system - the
hyperbola billiard. In section 5 we discuss the relevance of our
result for correlations in quantum spectra.

\section{Theory}

There are mainly three methods for the determination
of resonance spectra.
The pedestrian way is to numerically simulate the system,
compute the correlation function for some physical
observable(s), make a Fourier transform and then locate
its poles. This is of course very tedious and the success moderate and
the theoretical understanding gained minor.
We will thus focuse on
the remaining two methods.

\subsection{The zeta function}

Periodic orbit expansions (or cycle expansions - we will use the
words cycles, periodic orbits or simply orbits synonymously in the
following)  of {\em zeta functions} have shown successful for
calculating various averages of chaotic sets. Most of the labour has so
far
 been devoted
to maps, that is, systems with discrete time.
A zeta function for systems with continous time has been derived
in \cite{flows}. We briefly review the derivation.
The resonances $k_n$ may formally be associated with the eigenvalues
$exp(ik_n t)$ of the
evolution operator $\pounds ^t$, acting on a function $\Phi(x)$ as
\begin{equation}
\pounds ^t  \Phi(x)=\int \delta (x-f^t(y))\Phi(y)dy   \ \ .
\end{equation}

The phase space point $x$ is taken by the flow to $f^t(x)$ during time
$t$.
Our main concern in this paper is to compute the trace of this operator,
that is, the sum of its eigenvalues.
The trace may be written as a sum over the isolated
periodic orbits in the system
\begin{equation}
tr \pounds ^t \equiv tr(t) =\int \delta (x-f^t(x))dx=
\sum_p T_p \sum_{n=1}^{\infty} \frac{\delta(t-nT_p)}
{\Abs{det(1-M_p^n)}}  \ \ ,
\label{eqn:tracedef}
\end{equation}
where $n$ is the number of repetitions of primitive orbit $p$, having period
$T_{p}$ and  $M_{p}$ is the Jacobian of the Poincar\'{e} map.
This is exactly the sum occuring in
ref. \cite{Ozo} which is shown to approach unity as $t \rightarrow \infty$,
provided the system is bound (and no other system
will be discussed from now on).

The function $tr(t)$, (we will refer to this
object as simply {\em the trace} in the following)
may be written as the
Fourier transform of the logaritmic derivative
of a {\em zeta function}:
\begin{equation}
tr(t) = \frac{1}{2\pi i}
\int_{-\infty}^{\infty} e^{ikt}\frac{Z'(k)}{Z(k)}dk   \ \ .
\label{eqn:tracefour}
\end{equation}
We restrict ourselves to systems with two degree's of freedom
for which all periodic orbits are isolated and unstable.
The zeta function then
reads
\begin{equation}
     Z(k)=\prod_{p}\prod_{m=0}^{\infty}
          \left(1-\frac{e^{-ikT_{p}}}
    {\Abs{\Lambda_{p}} \Lambda_{p}^{m}}\right)^{m+1}
          \ \ ,                       \label{eqn:Zcl}
\end{equation}
where $\Lambda_{p}$ is the expanding
eigenvalue of $M_p$.
Each zero $k_{\alpha}$ of $Z(k)$ induce a pole
in the integrand and is identified with an eigenvalue of the evolution
operator. The leading zero is the escape rate, which is zero for a bound
system; $k_0=0$.

The corresponding zeta function for systems with a hyperbolic analytic
Poincar\'{e} map has been shown to be entire provided the existence of a
finite Markov partition \cite{Rugh}.
If the
zeta function is entire we can write
\begin{equation}
tr(t) =1+\sum_{\alpha \neq 0} e^{ik_{\alpha}t}
\label{eqn:ettplus}
\end{equation}
In eq. \EqRef{ettplus} we have neglected the integral
along the upper semicircle. It may be argued that this can only yield
a deltafunction $\delta(t)$ and derivatives of the deltafunction
$\delta^{n}(t)$.

If there is a gap between the real axis and the nontrivial zeros then
the trace approach unity exponentially fast.

A an example consider
the following simplified
Axiom-A system (it may be realized by a three disk scatterer with the
disks infinitely separated).  We take the system to have a binary complete
symbolic dynamics. An orbit $p$ has eigenvalue $\Lambda_p = \Lambda_0^{n_p}$
and period $T_p =n_pT_0$ where $n_p$ is the length of the symbol code $p$.
Expanding the product over cycles gives, c.f. ref. \cite{AAC}
\begin{equation}
Z(k)=\prod_{m=0}^{\infty}
          \left(1-2\frac{e^{-ikT_{0}}}
    {\Lambda_{0}^{m+1}}\right)^{m+1}  \ \ .
\end{equation}
The product is convergent so we can just read of the zeros
\begin{equation}\begin{array}{ll}
k_{n,m}=\frac{1}{T_0} \left( 2\pi n +i [ (m+1)\ln \Lambda_0 -\ln 2 ]
\right)  &
n=0,\pm 1 \pm 2 \ldots \ \ .
\end{array}\end{equation}
The spacing between zeros in the
real direction is directly related to the single timescale in the system -
the period of the shortest periodic orbit.
The situation for bound systems is generally much more complicated, as we
will see later; they may exhibit an infinite set of time scales.

For a more general system with complete binary grammar the spectrum
above will give the gross structure, whereas longer periodic orbit
provide corrections \cite{AAC,CE}. However, even for small deviations
from the simple example above there will be quite intricate cancellations
between poles and zeros \cite{Rugh,AAC,Russ} of the individual $m$-factors.
For bound systems there is no finite Markov partition
and no hyperbolicity. There is then no reason to expect entireness of the
zeta function and we will concentrate on revealing the
main singularities.

\subsection{The BER approximation - a probabalistic approach}

In ref. \cite{BER} a method has been given for the
determination of the approximate
resonance spectrum for intermittent systems.
In such a system there are two, more or less, distinct
phases; one regular and one irregular
(chaotic).
Call the consecutive instants when the system enters the regular phase
$\{ t_i \}$ and consider the intervals  $I_i=[t_{i-1},t_i]$.
Provided the chaotic phase is {\em chaotic enough},
the motions in different intervals are nearly mutually independent.
In particular the lengths of these intervals
$\Delta_i=t_i - t_{i-1}$ are mutually independent
and $\Delta$ may be considered as a stochastic variable
with probability distribution $p(\Delta)$.
Consider now the Fourier transform
\begin{equation}
\tilde{p}(k)=\int_{0}^{\infty}e^{-ik\Delta}p(\Delta)d\Delta \ \ .
\end{equation}
Under the assumption above,
the roots of
\begin{equation}
1-\tilde{p}(k)=0
\end{equation}
will now provide the approximate resonances.
We will modify the calculation of ref. \cite{BER} so as to apply directly to
the trace of the evolution operator. This will enable us to
find a very simple approximation of $Z(k)$ in terms of $p(\Delta)$.

First we turn the phase space integral
in eq. \EqRef{tracedef} into a time
average.
\begin{equation}
tr(t) = V \langle \delta (x(t_0)-x(t_0 +t)) \rangle
_{t_0} \ \ .
\end{equation}
where $V$ is the available phase space volume.

Let now $tr_m(t)$ be the trace conditioned by 'if $t_0 \in I_n$ then
$t+t_0 \in I_{n+m}$'. The trace is now expressed as the sum
$tr(t)=\sum_{m=0} tr_m(t)$.
We now define two probability distributions:
\begin{itemize}
\item $f_1(x,u)$ is the probability that the system is at
phase space point $x$ at time
 $t_0$ and that there is time $u$ left of the current interval.
\item $f_2(y,v)$ is the probability that the system is at point $y$ at time
 $t_0+t$ and that time $v$ has ellapsed since the current interval
 was entered.
\end{itemize}
We can now express the average as
\begin{equation}
tr_m(t) \approx V < \Delta >
\int du \; dv  \; p^{*(m-1)}(t-u-v) \int
dx \; dy \; \delta(x-y)f_1(x,u) f_2(y,v)  \ \ ,
\end{equation}
where $p^{*n}(t)$ is the n-fold convolution of $p(t)$, and $< \Delta >$
the mean length of the intervals.
We now express the convolution by means of the Fourier transforms
\begin{eqnarray}
tr_m(t)\approx \frac{ V < \Delta >}{2\pi} \int dk \;  e^{ikt}
\tilde{p}(k)^{m-1}  \int du \; dv  \;
dx \; dy \; \delta(x-y)e^{iku}f_1(x,u)e^{ikv} f_2(y,v)= \nonumber  \\
\frac{ V < \Delta >}{2\pi} \int  dk\; e^{ikt}\tilde{p}(k)^{m-1} \int dx
\tilde{f}_1(x,k) \tilde{f}_2(x,k)   \ \ .
\end{eqnarray}
We can now sum over $m$ and arrive at
\begin{equation}
tr(t)\approx tr_0(t)+\frac{ V < \Delta >}{2\pi} \int  dk\; e^{ikt}
\frac{1}{1-\tilde{p}(k)} \cdot
 \int dx
\tilde{f}_1(x,k) \tilde{f}_2(x,k)  \ \ .  \label{eqn:traceBER}
\end{equation}
Comparing with eq. \EqRef{tracefour} we indentify
\begin{equation}
\frac{Z'(k)}{Z(k)} \approx \frac{V < \Delta >i}{1-\tilde{p}(k)} \cdot
 \int dx
\tilde{f}_1(x,k) \tilde{f}_2(x,k)  \ \ .
\end{equation}
Moreover, we can identify:
\begin{equation}
Z(k) \approx \hat{Z}(k) \equiv 1-\tilde{p}(k)   \label{eqn:hatZdef}
\end{equation}
provided that
\begin{equation}
V < \Delta > \int dx
\tilde{f}_1(x,k) \tilde{f}_2(x,k) =
\int \Delta e^{-ik\Delta}p(\Delta)d\Delta  \ \ .  \label{eqn:verif}
\end{equation}
To verify that this is indeed the case
we make use of the fact the remaining
time of the current interval $u$ is a function of the phase space point
$u=g_1(x)$ so that \begin{equation}
f_1(x,u)=\phi(x)\delta(u-g_1(x))
\end{equation}
where $\phi(x)$ is the phase space density.
Similarly we have
\begin{equation}
f_2(x,v)=\phi(x)\delta(v-g_2(x))  \  \  .
\end{equation}
We can now express the probability function $p(\Delta)$ weighted with the
length of the interval $\Delta$ in terms of $g_1$ and $g_2$
\begin{equation}
\Delta \cdot p(\Delta) = < \Delta > \int \phi (x)
\delta(\Delta-g_1(x)-g_2(x)) dx
\end{equation}
Also using the fact that the phase space density is uniform:
$\phi(x)\cdot V=1$ eq. \EqRef{verif} is easily verified.

\vspace{0.5cm}

In this section we have been able to give the zeta function a probabalistic
interpretation and have constructed an approximate zeta function $\hat{Z}(k)$
whose zeros should be interpreted as the approximate resonances. Both
functions, $Z(k)$ and $\hat{Z}(k)$ have the trivial zero $k_0=0$ for
unitarity reasons.

\section{Phenomenology of the BER approximation}

The close relationship between $Z(k)$ and $\hat{Z}(k)$ as defined by
eq. \EqRef{hatZdef} motivates a study of how different features of the
probability function
$p(\Delta)$ affects the spectrum and
trace of the evolution operator. Throughout this section we will ignore
the difference between $Z(k)$ and $\hat{Z}(k)$,
that is we will temporarly assume that $Z(k)=\hat{Z}(k)$.
The reader should keep in
mind that the spectra and traces obtained here is of course only
approximations of those of real dynamical systems and the errors are
not under control.

\vspace{0.5cm}

We begin by studying a model (model A) with a sharp
step and and exponential decay
\begin{equation}
p_A(\Delta)=\left\{
\begin{array}{ll}
0 & \Delta < T_{1} \\
N e^{-\Delta / T_2} & \Delta \geq T_{1}  \ \ , \end{array}\right.
\label{eqn:modelA}
\end{equation}
where $N$ is a normalization factor.
We take the Fourier transform and get
\begin{equation}
{Z}_A(k)=\frac{e^{-ikT_1 }}{1+ikT_2 }-1  \ \ .
\end{equation}
One can obtain the following asymptotic expression for the zeros
of this function
\begin{equation}
\begin{array}{ll}
k_n \approx \frac{1}{T_1} \left(
\pm 2\pi (n-\frac{1}{4})+i \cdot \ln (\frac{T_2}{T
_1}2\pi
(n-\frac{1}{4})) \right) & n=1,2,3,\ldots  \ \ ,
\label{eqn:Aspec}
\end{array}
\end{equation}
which agrees very well for large $n$ and/or large $\frac{T_2}{T_1}$,
c.f. fig. 1a.
We see that the relevant time scale is $T_1$, whereas $T_2$ is of minor
significance.
${Z}_A(k)$ has a simple pole at $k=i/T_2$
reflecting the behaviour of $p_A(\Delta)$ at infinity.
This is all we need to calculate the trace by means of
eq. \EqRef{tracefour}. We get
\begin{equation}
tr(t)=\Psi_{zeros}(t)+\Psi_{tail}(t)  \ \ ,
\end{equation}
where
\begin{equation}
\Psi_{zeros}(t)=
1+\sum_{n} e^{ik_{n}t} \approx
1+(\frac{T_2}{T_1})^{-t/T_1}\cdot
\sum_{n=1}^{\infty}2\cos (2\pi (n-\frac{1}{4})\frac{t}{T_1})\cdot
(2\pi (n-\frac{1}{4}))^{-t/T_1}  \ \ ,
\label{eqn:Asum}
\end{equation}
and
\begin{equation}
\Psi_{tail}(t)=-e^{-\frac{t}{T_2}} \ \ .
\end{equation}

The sum \EqRef{Asum} is absolutely convergent when $\frac{t}{T_1}>1$
and tends
exponentially to unity as $\frac{t}{T_1} \rightarrow \infty$.
To study its behaviour for small $\frac{t}{T_1}$ we have to the
smear the function term by term, that is convolute it with a gaussian
having
width $\sigma$. For a general spectrum $k_n =\pm a_n + i \cdot
b_n$ we find for
the smeared sum
\begin{equation}
\Psi_{zeros,\sigma}(t)=
2\sum_n \exp \left(-\frac{\sigma^2}{2}(a_n^2-b_n^2)-b_n
t\right) cos(a_n t-\sigma^{2}a_n b_n)  \ \ .
\label{eqn:smear}
\end{equation}
For all relevant spectra this sum is absolutely
convergent for any smearing width $\sigma$.

The trace when $\frac{T_2}{T_1}=0.5$, is plotted
in fig. 1b. The sharp step at $\frac{t}{T_1}=1$ reflects the sharp step in
$p_A(\Delta)$. The are indeed steps whenever $\frac{t}{T_1}$
is an integer, but due to the exponential damping they are less
pronounced.

\vspace{0.5cm}

Let us now consider a case with power law decay of $p(\Delta)$ (model B).
\begin{equation}
p_B(\Delta)= \left\{
\begin{array}{ll}
0 & \Delta < T_1 \\
N (\frac{T_1 }{\Delta })^{m} & \Delta \geq T_1 \ \ . \end{array}
\right.
\label{eqn:modelB}
\end{equation}
Again we compute the Fourier transform and seek the zeros of
\begin{equation}
{Z}_B(k)=(m-1)\cdot E_{m}(ikT_1 )-1  \ \ ,
\label{eqn:ZB}
\end{equation}
where $E_{m}(z)$ is the exponential integral.
It turns out that all zeros a positioned where we can use the asymptotic
expansion of $E_{m}(z)$, see \cite{Abra}.
The position of the
zeros are well approximated by
\begin{equation}
\begin{array}{ll}
k_n \approx \frac{1}{T_1} \left( \pm 2\pi (n-\frac{1}{4})+
i \cdot \ln (\frac{2\pi
(n-\frac{1}{4})}{m-1}) \right) & n=1,2,3,\ldots  \ \ ,
\end{array}
\label{eqn:Bspec}
\end{equation}
which is equivalent with {\em model A} when $m-1=\frac{T_1}{T_2}$, that is
when the height of the step in $p(\Delta)$ is the same.
High zeros (large $n$) are thus only sensitive to the immediate vicinity of
the step. In fig. 1a the exact zeros of model A and B are compared
with this asymptotic result.
 Formally, if we increase $m$ (or $\frac{T_1}{T_2}$ in model A)
${p}_B(\Delta)$ will tend to a delta peak,
and the resonances approach the real axis, resembling an integrable system.
Of course, the assumption in the
BER approximation is then no longer fulfilled.

However there is one major difference to model A.
The zeta function
is no longer meromorphic since $E_{m}(ikT_1)$ has a branch cut along the
positive imaginary $k-$axis.
This is  a consequence of the power law tail of $p_B(\Delta)$.

We therefore have to modify the contour used in
section $2.1$
\begin{equation}
tr(t) = \frac{1}{2\pi i}
\int_{\gamma_{branch}} e^{ikt}\frac{Z'(k)}{Z(k)}dk+
\sum_{\alpha} e^{ik_{\alpha}t} \ \ ,
\label{eqn:grenlap}
\end{equation}
where $\gamma_{branch}$ is given in fig. 2.
We now get
\begin{equation}
tr(t)= \Psi_{zeros}(t)+\Psi_{tail}(t) \ \ ,
\end{equation}
with
\begin{equation}
\Psi_{tail}(t)= \int_{0}^{\infty}\frac{1}{\pi} e^{-\xi x}
Im (\frac{Z'(i\xi + \epsilon)}{Z (i\xi + \epsilon)})d\xi \ \ ,
\end{equation}
and $\Psi_{zeros}(t)$ as before.
$\epsilon$ is a small positive number.
The resulting trace for model B is shown in fig. 1b.
The discrete  $\Psi_{zeros}(t)$
is the same as in the previous model (A) since we are using the asymptotic
expressions \EqRef{Aspec} and \EqRef{Bspec}.
It is straight forward to show that $\Psi_{tail}(t) \rightarrow -1;
\; \; t \rightarrow 0$ and $\Psi_{tail}(t) \rightarrow
-\frac{T_1}{2t}; \; \; t \rightarrow \infty$. The asymptotic behaviour of
the trace for general $p(\Delta )$ with power
law tail is given at the end of
this section.

\vspace{0.5cm}

Let us modify the first model (A) so as to have a
$C^1$ discontinuity instead of a $C^0$.
\begin{equation}
p_C(\Delta)= \left\{
\begin{array}{ll}
N & \Delta< T_{1} \\
N \exp (\frac{T_{1}-\Delta}{T_2}) & \Delta \geq T_{1} \ \ . \end{array}
\right.
\label{eqn:modelC}
\end{equation}
There is again a pole at $k=i/T_2$.
The zeros are now approximated by
\begin{equation}
\begin{array}{ll}
k_n \approx \frac{1}{T_1} \left( \pm 2\pi n+
2i \ln(\frac{T_2}{T_1}2\pi n) \right) & n=1,2,3,\ldots \ \ .
\end{array}
\end{equation}
Apart from shift a in the real direction, as compared to model A (and B),
 the important effect is
the factor $2$ in the imaginary part - the zeros
lie further away from the real axis. The corresponding trace is given in
fig. 3.

%

\vspace{0.5cm}

Finally let us demonstrate how small irregularities in $p(\Delta)$ will have
a drastic effect on the spectrum. We will consider
\begin{equation}p_D(\Delta)=\left\{ \begin{array}{ll}
0 & \Delta< T \\
N(\Delta -T)\exp (-\frac{\Delta}{T}) & T \leq \Delta < 2T \\
N(\Delta -T)\exp (-\frac{\Delta}{T})(1+\epsilon) & 2T \leq \Delta \ \ .
\end{array}
\right.
\label{eqn:modelD}
\end{equation}
${Z}_D(k)$ has a double pole at $k=i/T$.
The spectrum of zeros for different perturbations, $\epsilon$, is given in
fig. 4b. In the low part the unperturbed spectrum is only sligthly affected.
The perturbation corresponds to a nonleading generations of resonances.
But eventually the two generations interfere, leading to
a modulation in the spectrum. For high $t$
the higher time scale (2T) takes over and
dominates.
This result leaves us with considerable doubt on the possibility on
determining
the resonance spectrum from a numerically obtained probability distribution
$p(\Delta)$. If there is a single dominating time scale it seems perfectly
feasible \cite{BER}. But there might hide further time scales,
with considerable potential impact on the spectrum.
The traces for $\epsilon=0$ and $\epsilon=0.01$
are given in fig. 4c. For the unperturbed case, the trace is essentially
zero when $0<\Delta < T$. This is a remarkable conspiracy
between $\Psi_{zeros}$
and $\Psi_{tail}=-2\cdot e^{-t/T}$ to cancel in this way. But from a physical
point of view it seems sensible, c.f. eq. \EqRef{tracedef}. There is,
in any
system, a shortest periodic orbit, with period $T_{min}$, such that
$tr(t)=0$ when $t<T_{min}$.

When we add a perturbation, $\epsilon$, the trace changes in two respects.
First,
$p_D(\Delta )$ has a small step at $t=2T$, which is not visible
in the figure. When $0<\Delta < T$ the trace is no longer positive definite.
This is not to allarming since we have neglected $tr_0(t)$ in
eq. \EqRef{traceBER}. Moreover, the spectral sum is divergent in this region
so it is questionable what confidence we should have in the results there.
For larger $t$ we see that
a small change of $p(\Delta )$ corresponds a small change in
the trace.

 \vspace{0.5cm}

Let us summarize our findings so far, remembering that our evidence is
only heuristical and that we have assumed the exact equality
$Z(k)=\hat{Z}(k)$. The zeros of the zeta function depends almost entirely on
the breakpoints  (i.e. different kinds of discontinuities)
in the
probability distribution $p(\Delta)$. Each such breakpoint
introduces a time
scale. When several such timescales are present the largest will dominate
sufficiently high up in the spectrum, in the sense that the spacing between
succesive zeros in the real direction is directly related to this timescale.

The zeta function is entire only if $p(\Delta)$ decays faster
than exponentially. If the decay is exponential the zeta function will have
a pole on the imaginary axis. In this case the trace will approach unity
exponentially fast.
 If $p(\Delta)$ decays as a powerlaw there will be a
logaritmic branch cut along the positive imaginary axis implying powerlaw
convergence of the trace. It is straightforward to show that if
$p(\Delta ) \sim 1/\Delta^m$, the tail
contribution, $\Psi_{tail}(t)$, to the trace, will behave asymptotically as
\begin{equation}
\Psi_{tail}(t) \rightarrow
\frac{\Delta_0 ^m p(\Delta_0)}{(m-1)< \Delta >}\frac{1}{t^{m-1}} \; \; ;
\; \;  t \rightarrow
\infty
\end{equation}
with $m\geq 3$,
where $< >$ denotes mean value.
$\Delta_0$ is any point in the tail of $p(\Delta)$.

\section{Cycle expansions - studies of a dynamical model}

In this section
we will study periodic orbit expansions
of the zeta function \EqRef{Zcl} for a specific dynamical
system, namely
the
{\em hyperbola billiard}, i.e., a point particle
elastically bouncing off the walls given by the equation
\begin{equation}
     \Abs{xy}=1  \ \ ,                            \label{eqn:hyp}
\end{equation}
An effective algoritm exist for the computation of the
cycles
in this system
\cite{SS}. For a definition and discussion
 of the symbolic coding of cycles used in
this paper, please consult ref. \cite{DR2}.
We use units such that the particle mass $m=1$ and we fix the energy
to $E=\frac{1}{2}$, the length and time scales thus coincide.
The zeta function may be factorized into the irreducible representation of
the symmetry group \cite{CEsym}, which is
$C_{4v}$ for the hyperbola billiard.
We will restrict our attention to the symmetric representation $A_1$
containing the {\em trivial} zero $k_0=0$, \cite{CEsym}.
This is equivalent to
restricting the system to the fundamental domain \cite{DR2,CEsym}.
The periodic orbits considered
in the following are periodic in this fundamental domain. They have to be
retraced 1,2 or 4 times to close onto themselves in the full system.
Although we will frequently draw pictures of the
periodic orbits in the full system since
they are easier to visualize that way.

First we expand the $m$-product of \EqRef{Zcl}
\begin{equation}
       Z(E)=\prod_{p} \prod_{m=0}^{\infty}
 \left(1-\frac{e^{-ikT_{p}}}
    {\Abs{\Lambda_{p}} \Lambda_{p}^{m}}\right)^{m+1}=
   \prod_{p} \sum_{n_{p}=0}^{\infty} b_{n_{p}}(\Lambda_p)(-1)^{n_p}
\frac{e^{-ikn_pT_p}}{\Abs{\Lambda_p}^{n_p}}  \ \ .
\end{equation}
Reccurence relations for the expansion coefficients $b_n$
are derived in the
appendix.
If we now expand the $p$-product,
we obtain a Dirichlet series consisting of all distinct
linear combinations (pseudo orbits) $N=[n_{p}]$,:
\begin{equation}
       Z(E)=\sum_{N}C_{N}e^{-iT_{n}}
       \ \ , \label{eqn:Zn}
\end{equation}
where we have defined the quantities
\begin{equation}
        C_{N}=\prod_p
\frac{(-1)^{n_p}b_{n_{p}}(\Lambda_p)}{\Abs{\Lambda_p}^{n_{p}}}  \ \ ,
\end{equation}
\begin{equation}
        T_{N}=\sum_{p}n_{p}T_{p}  \ \ .
\end{equation}

There are no
obvious choice of how to order the terms, but natural choices are to order
them according to i) length of the symbol code,
 ii) increasing period $T_N$
or,  iii) decreasing {\em amplitudes} $C_N$.
Our default option will be the last one.
Switching between the three options will
mean a complete reorganisation of the sequence the pseudo orbits.
Our default choice means that we can make no
simple statements about the conditional convergence of
the expansion.  This will however not bother us to much. Typically,
nontrivial zeros of zeta functions are positioned where
the products, or series, are not even conditionally
convergent, c.f. the Riemann zeta function, although in some lucky cases,
due to alternating signs of the $C_N$'s, they will lie in the strip
of conditional convergence \cite{Aur}.


The idea is now to
compute the partial sums (the first $N_{cut}$ terms)
and determine the
zeros of the truncated expansion
and study their possible convergence
when $N_{cut}$ is increased. To this end we have computed all prime
orbits with eigenvalue $\Lambda_p<2700$, and all orbits with $T<14$, both
samples containing about $10^3$ orbits.
 To have something to crosscheck our result with, we
calculate the trace by means of eq. \EqRef{tracedef}. Of course, we have to
use smearing so we replace the delta function in eq. \EqRef{tracedef} with a
gaussian having width $\sigma =0.5$ The result is given in fig. 10.
Due to the exponential proliferation of cycles this method
soon becomes completely
unrealistic as $t$ is increased.

Our first attempt is to simply include all orbits up to the cutoff.
 The zeros thus found will
show no tendency of convergence (the zeros drift monotonously towards the
origin as $N_{cut}$ is increased),
and the trace, computed by means of the sum
over zeros (again using smearing, c.f. eq \EqRef{smear}),
will not show the slightest resemblance with the exact trace.
Such failure of
convergence are usually due to some singularity in the zeta function.
In order to identify this
we will, in some detail, study the structure of the set of cycles.
Inspired
by the BER approximation we
divide (somewhat ambigously) the billiard into one chaotic part in
the center and one (almost)
regular out in the arms. We can thus divide the cycles
into three subsets.
\begin{enumerate}

\item Cycles with bounces exclusively in the chaotic region. The eigenvalues
$\Lambda$ are exponentially bounded with length.

\item Cycles with bounces in both regions.

\item Cycles with  bounces exclusively in the regular arms.

\end{enumerate}

In the following we will focuse our attention on the types of cycles which
dominate the expansion, with ordering chosen as above.
These are the pseudo orbits with large $C_N$ which means that the
important cycles come from set
2 and 3 above - cycles spending most of the time in the arms.

It is also convenient to distinguish between
{\em simple} cycles, that is cycles making only
one (or no) excursion into an arm, and {\em composed}, that is
cycles making more than one such
excursion. The simple cycles occur in sequences
of the form $p0^j$ where the row of
$0$'s corresponds to a number ($\sim j$) of
oscillating bounces in an arm and $p$ corresponds to the
behaviour in the central region.

Some examples of simple
cycles are given in the table
and fig. 5. Cycles belonging to the sequences $20^j$ and $110^j$ make
one excursion into an arm and are
then injected into the opposite arm without any bouncing in the
central chaotic region.
 These two sequences are in fact unbounded.
All other simple sequences are, we believe,
finite.
A cycle $210^j$ makes a
number of oscillations in one arm, sneaks into
a neighbouring arm without bouncing in the middle. This is obviously a very
short sequence.
A cycle in the sequence $220^j$ makes a number of
oscillations in one arm, makes one
scattering in the central region and is then
injected back into the arm it came from.
A cycle $2210^j$ makes a
number of oscillations in one arm, bounces once
in the central region and is then
injected into a neighbouring arm.

It is obvious from fig. 5 why finite
families
$p0^j$ are terminated at some $j=j_{max}$. There is a section of the
orbit in the central region coming closer and closer
to a branch of the boundary as
$j$ is increased. When $j=j_{max}+1$ the orbit would need to go through this
branch which it of course is not allowed to. For this reason the finite
sequences are naturally organised into families where one member go directly
to the opposite branch when another member hit a branch on the way, see fig.
5a .  Thus $210^j$ may be considered as a member
of $(210^j,10^{j+3})$.
Other important familes are $(220^j,2110^j,1210^j,1010^{j+1})$ and
$(2210^j,2010^{j+1},120^{j+1},1110^{j+2})$.
All members of a family
terminate at the  same $j=j_{max}$,  cf. refs \cite{DR2,Kai}.
The family member with the fewest bounces in the central region has the
largest weight and the rest are of minor importance.

The class consisting of infinite simple
sequences, will
be named ${\cal C}_{inf}$ in the following and contains only $20^j$ and
$110^j$.
The class of simple
finite sequences is named ${\cal C}_{fin}$. The orbits in these
classes are taken from
sets 2) and 3) above respectively. The orbits in subset 1) may be
absorbed into e.g. class ${\cal C}_{fin}$.

Composed cycles doing two excursions into the potential arms may be written
on the form $p_1 0^{j_1}p_2 0^{j_2}$.
Picking the simple cycles (or rather symbol strings) $p_1 0^{j_1}$ and
$p_2 0^{j_2}$ from ${\cal C}_{inf}$ and/or ${\cal C}_{fin}$
we can construct three
classes  of such composed cycles:
${\cal C}_{inf}*{\cal C}_{inf}$, ${\cal C}_{inf}*{\cal C}_{fin}$ and ${\cal
C}_{fin}*{\cal C}_{fin}$.
Generalisation to higher orders are straightforward.
Members of the 2-composed sequences $20^{j_1}20^{j_2}$,
$20^{j_1}210^{j_2}$ and $210^{j_1}210^{j_2}$ (from the three classes
respectively) are displayed in fig. 6. The latter two are again finite. The
allowed cycles may be represented in the $j_1,j_2$ plane as in fig. 7.

The region of allowed cycles is quite different for the class
${\cal C}_{inf}*{\cal C}_{inf}$. Take for example the
$20^{j_1}20^{j_2}$ sequence. In the
adiabatic approximation one can show that, in the limit $j_1 \gg 1$ and
$j_2 \gg 1$, the allowed cycles lie in the infinite region
\begin{equation}
\frac{j_1}{a} < j_2 < a\cdot j_1  \ \ ,
\end{equation}
where $a \approx 1158$.

Of course, the division between simple and composed cycles is
again somewhat ambigous;
e.g. should $21210^j$ be considered as simple or to belong
to the composed sequence $210^j_1210^j_2$, c.f. fig. 5b and 6a?

Let us now discuss the contributions of the different sequences
to the expansion of the zeta function.

We begin with the class ${\cal C}_{inf}$.
For large $j$ it was shown in
\cite{PD1} that the periods and the stability
eigenvalues for the $20^j$ and $110^j$ sequences are:
\begin{equation}\begin{array}{ll}
T_j \approx \sqrt{4\pi (j+1)}  & \\
\Lambda_j \approx c(j+1)  & j=1,2,\ldots ,\infty \ \ ,
\end{array}\end{equation}
where $c=117/4$ for $20^j$. The corresponding factor for $110^j$ is much
bigger.
We suspect these infinitite sequences to be the main cause of
our divergence  problem. Before following this track we discuss the finite
sequences.

A finite simple sequence $p0^j$ will have invariants
depending on $j$ similarly to the infinite ones above, but with $j$
terminating at some $j_{max}$, see the table. The slope of $\Lambda_j$
versus $j$  will depend on the behaviour in the
central region and seems to be exponentially bounded with the number of
bounces there. Each finite
sequence naturally introduces a time scale given by the period of the last
orbit, some of these timescales are named in the table.
The trace, fig 10, will obviously show a sudden
decrease, cf. eq. \EqRef{tracedef} at any such time scale. In fig. 10 they
are rather recognized as smooth peaks due to the smearing.
These (infinite) set of timescales are essential for the problem and
will be reflected by the zeros of the zeta function.

The crossterms between simple orbits are naturally discussed on the same
footing as the composed sequences. There is a potential possibility of
shadowing, that is, the term due to orbit $p_1 0^{j_1}p_2 0^{j_2}$ is
compensated by the pseudo orbit consisting of $p_1 0^{j_1}$ and
$p_2 0^{j_2}$. This can however only be partly realised due to the
finiteness ({\em pruning}) of the sequences.
The double hatched area in fig 7a shows where there exist compensating
crossterms.
However  these terms are
sparse in the expansion, and may in some sense be considered as corrections,
c.f. the {\em curvature corrections} in ref. \cite{AAC}.
The orbits along the boundary of the allowed region, c.f. fig. 7, also
introduce timescales.

We suspected the class ${\cal C}_{inf}$ to be the main cause of our problem.
We will now attempt a factorization
\begin{equation}
Z(k)=Z_1(k)\cdot Z_2(k)  \ \ ,
\end{equation}
where, in the first factor, we only include cycles $p \in {\cal C}_{inf}$ and
in the second we include the rest $p \in {\cal C}_{tot}-{\cal
C}_{inf}$. The first factor is approximately
\begin{equation}
Z_{1}(k)\approx
1-\frac{1}{c}\sum_{j=1}^{\infty}\frac{e^{-i\sqrt{4\pi j}k}}{j} \ \ .
\end{equation}
Crossterms are sparse in the series, so we have omitted them. It is
straight forward to see that the function defined by this sum have a
logaritmic branch cut
along the positive imaginary axis. 

Our hope is now that the eigenvalues are given by the zeros of $Z_2(k)$,
which is expanded as before.
We now seem to obtain a well defined
spectrum where when $N_{cut}>180$ see fig. 8!
However, we do not find
much
improvement when $N_{cut}$ is increased further.
It is thus obvious that
the zeta function needs further regularization.
The convergence of the trivial zero $k_0$ is plotted in fig. 9.

We obviously need some support for our claim that the obtained spectrum
is indeed close to the exact one. To that end
we compute the trace by means of the sum over zeros $\Psi_{zeros}(t)=
1+\sum_{n} e^{ik_{n}t}$ and compare with the exact result due to eq.
\EqRef{tracedef}.
Again we use gaussian
smearing, which is unavoidable since we only have a finite sample of zeros of
$Z(k)$. The result is plotted in fig. 10. We see that
the curves closely resembles each other on the small scale structure.

Let us first study the structure of the spectrum, cf. fig. 8.
We recognize several features from the spectra
in section 3.
The imaginary part of the zeros increase slower than logaritmically with
the real parts (cf. model D in section 3).
We also recognize a modulation of
the spectrum. These features indicate that several time scales
are involved.
The real part
of the zeros are more
or less equally spaced. The mean spacing correspond to a time scale
$T\approx 27.8$ which is much longer than the shortest periodic orbit
of the system, $T_{min}=2$. It is indeed close to the timescale
$T_c\approx 27.61$
associated with the sequence $220^j$ which dominates the expansion of $Z_2$
for the first 180 terms. Clearly, using ordering option ii) we would need
all cycles with $T_p \leq T_c$ in order to resolve the spectrum.
This would mean $\sim 10^6$ cycles. With the ordering option chosen we have
managed to obtain a good approximation to the spectrum using a thousand times
fewer orbits!

In fig. 10 we see that the exact trace and the sum $\Psi _{zeros}$
differ by a slowly varying
function. The immediate suggestion to
explain this discrepancy is that there is
another contribution to the trace from a
pole or a branch cut, c.f. $\Psi_{tail}$.
The latter possibility is suggested by
the function
$Z_1(k)$ having a branch cut.
However we do not yet claim that $Z_2(k)$, as defined above, is entire.
We cannot expect the branch cut to be factored out that easy. This means
that we cannot compute this extra contribution only from $Z_1$.

The problems encountered so far
could be somewhat enlightened by explicit knowledge of
the probability distribution $p(\Delta)$ for the billiard since we expect
there to be a close relationship between its Fourier transform
and the zeta function.
This is easily obtained by simulating
the system. It is thus essential that we take advantage of
the adiabatic invariance in the arms, c.f. \cite{PD1}.
There is no sharp borderline between the chaotic and regular part of the
billiard. We make an arbitrary division at some $x=x_{div}$ (remember the
system is defined on the fundamental domain $y>0,y<x$).
The result is given in fig. 11. The histogram is based on $10^6$ visits
in the arm and the borderline
is set to
$x_{div}=4$.

We recognize the timescales as introduced above and defined
in the table, as underlying the structure of $p(\Delta)$.

For small $\Delta<15$
the result is sensitive to the exact location of the borderline.
But the rest of $p(\Delta)$  is rather insensitive to it.

The curve exhibit
a peak around $\Delta \approx T_b$. Among the {\em important} composed
sequences $210^{i_1}210^{i_2}$ and $20^{j_1}210^{j_2}$ (with $j_1 \neq 0$)
all cycles  have $T< T_c$, and indeed their periods are distributed around
the peak of $p(\Delta)$.

When $T_c < \Delta < T_d$, $p(\Delta)$ is
decreasing exponentially to a high degree of accuracy.
In the region of exponential decay there seems to be
a gap, with no pronounced timescales (as defined by means of finite
sequences).

When $\Delta > T_d$, the curve slowly begins to deviate from the exponential.
Relevant sequences for this tail are e.g.
$20^{j_1}20^{j_2}$ -
the sequence $2020^j$ terminates at $j \approx 990$, $T \approx 116$.

It would
be highly desirable to know the
asymptotic
form of $p(\Delta)$, if possible we could be able to estimate the
branch cut contribution $\Psi_{tail}(t)$.
However, the crossover from exponential decay to something else
(powerlaw?) warns us
that
$p(\Delta)$ is not easily extrapolated. There may
hide more surprises higher up.
Estimating the tail behaviour by analytical means
would require a detailed understanding of how the
almost integrable arms are coupled through the chaotic central region.
The problem lies in the
very long laminar phases (large $\Delta$) which have segments of the
trajectory
coming from
one arm and injected directly into the opposite without bouncing in the
chaotic part (in the simulation we require at least one central bounce
to consider the laminar phase as terminated). From a cycling point
of view
the
classes ${\cal C}_{inf}$, ${\cal C}_{inf}*{\cal C}_{inf}$ etc. are
central for the description of these trajectory. Excluding ${\cal C}_{inf}$
as a first step in the
regularization, as we did, seems quite reasonable.
However, to proceed the regularization scheme further will
require entirely new
ideas. Such a procedure would probably reveal more
singularities.

In the calculation of the spectra above we concluded that the timescale
$T_c$ is in some sense dominant. An explanation of this is offered by
the shape of $p(\Delta)$. $T_c$ is the last timescale (breakpoint)
before the
exponential part of $p(\Delta)$. We conlcuded in section 3 that such a
behaviour of $p(\Delta)$ would lead to a mean spacing corresponding to
$T_c$, at least in the lower part of the spectrum. We stress again that
it is not in practice possible to calculate the spectrum from $p(\Delta)$
obtained by numerical simulation
since it is the fine structure of $p(\Delta)$ that determines the spectrum
and this may not be resolved by any reasonable statistics.

\vspace{0.5cm}

Allthough this section ended in some frustration we summarize our main
achievements: We have computed several isolated zeros of the zeta function
for the hyperbola billard. We have also presented evidence for a branch cut
in the zeta function implying slower-than-exponential decay of the trace.
Finally we have identified a dominant timescale which is much longer
than the shortest periodic orbit of the system.

\section{Concluding remarks}

In this paper we have discussed relations between
periodic orbits, the resonance spectrum and the
dynamical behaviour of a system. We have suggested a close
relationship between the
classical zeta function $Z(k)$ and the corresponding function $\hat{Z}(k)$
as obtained from the BER approximation. We have not pursued this relation
very far for the hyperbola billiard. This system is peculiar having infinite
phase space volume so that the derivation in section 2.2 becomes dubious.
To study these ideas
in detail will require much more theoretical work as well as
numerical studies of a range of systems.
Ergodic billiards calling to be investigated in
this respect are e.g. the Sinai
billiard and the closed three disk billiard. For the Sinai
billiard it is straightforward
to show that the function $p(\Delta)$ should decay as $1/\Delta ^{3}$
suggesting that the trace behaves asymptotically behaves as $tr(t)
\rightarrow  1-const/t$.
Note that for
this systems eq. \EqRef{tracedef} is not valid; the marginally stable orbit
causes problems.

For the case of the hyperbola billiard we observed that the relevant times
scale  was much larger than
the period of the shortest cycle.
The dominant timescale for the Sinai billiard should be
the mean free path between the disks bounces which is also much longer than
the shortest cycle, for small disk radii.

So far we have only been discussing billiards.
The resctriction to billiards is not only a convenient one
(periodic orbits are much more easily found in billiards). Smooth systems
seem to differ from billiards
even qualitatively \cite{DR2,PD1,DR1}. The reason is that
ergodicity seems to be lost at the momemt the billiard
walls are made soft. To model this, it is convenient to embed the
billiard into a family of smooth Hamiltonians, e.g. the hyperbola billiard
is obtained as the $a \rightarrow 0$ limit of the family of potentials
$V(x,y)=\frac{1}{2}(x^2y^2)^{1/a}$. When $a$ is increased (the walls get
softer) the finite sequences terminate earlier and earlier, indeed the
sequences get considerably shorter when $a=1$ as compared to the
billiard case $a=0$ \cite{PD1}.
Orbits are
pruned off familywise (c.f. the
definition of family in section 4) through bifurcations. In ref. \cite{DR2}
it is demonstrated how the bifurcation of a family belonging to a {\em
simple} sequence also involve {\em composed} orbits. In each such bifuraction
cascade, some orbits get stable and ergodicity is lost. It does not seem
sensible that there can be any parameter value $a$ where no such
stabilization occurs. The occurence of stable islands seems to be a
general feature of smooth potentials.

A very interesting application of the {\em trace}
concerns correlations in quantum spectra.
The reason for this coupling is the close resemblance between the classical
zeta function \EqRef{Zcl} and the semiclassical one \cite{Vor}
\begin{equation}
     Z_{sc}(E)=\prod_{p}\prod_{m=0}^{\infty}
          \left(1-\frac{e^{iS_{p}(E)-\mu_{p}\frac{\pi}{2}}}
    {\Abs{\Lambda_{p}}^{1/2} \Lambda_{p}^{m}}\right)
          \ \ .                       \label{eqn:Zsc}
\end{equation}
The $\mu_p$'s are the Maslov indices and $S_p$ are the action integrals
which for homogenous potentials, such as billiards, are proportional to
the periods $T_p$ times some power of the energy $E$.
The zeros of this object are (in the semiclassical limit) the quantum
eigenvalues. The spectral density, as given by the logarithmic derivative of
$Z_{sc}$, yields the Gutzwiller trace formula \cite{Gut0}.

One such interesting correlation measure is the spectral
rigidity $\Delta_3 (L)$. It is defined as follows. Given the quantum spectra
$\{ E_i \} $, consider the spectral staircase function $N(E)=\sum_{i}
\theta (E-E_i) $. Now, make the best fit, by means of the least square method,
by a linear function over $L$ mean level spacings. Calculate the square
deviation between $N(E)$ and this linear function, average this quantity
over an energy interval $\Delta E$ which contains many levels but is
semiclassically small ($\Delta E$ is smaller than the energy scale given by the
shortest periodic obit in the system), and you end up with the function
 $\Delta_3 (L)$. Under the assumption that the trace, $tr(t)$, is exactly one
when $t > T_{min}$ Berry \cite{Berr}
showed that  $\Delta_3 (L)$ is consistent
with GOE for a range $0<L<L_{max}$ where $L_{max}$ corresponds to
$T_{min}$, if the system is chaotic and time reversible.
 But for bound chaotic systems we have realized
that the trace may be significantly different from one up to a timescale much
bigger that  $T_{min}$. This means that there is a region in $L<L_{max}$
where $\Delta_3 (L)$ {\em cannot} exhibit universal behaviour and where we
{\em can} perform semiclassical calculations. This is really what
{\em quantum chaology} is about, to relate the quantum behaviour to the
classical by semiclassical methods. Indeed, such departure from universality
for
 $L<L_{max}$ has been observed in a number of numerical calculations
\cite{Diet,Arve,Mart}.
Moreover, if there is a contribution from a branch cut of the zeta function
we expect the trace $tr(t)$ not to approach unity exponentially but rather
as a powerlaw. This extends further the dynamical
interesting region in $L$. Indeed one can show that if the trace goes
asymptotically as $tr(t) \rightarrow 1-c/t$ the universal (GOE) result can
only be achieved in the deep asymptotic limit $E \rightarrow \infty$.

Time scales in connection with the spectral rigidity has been discussed in
ref. \cite{Arve}.

We have seen that the trace for the hyperbola
billiard is closer to 0.5 than 1.0, for a considerable
range in $t$ .
This means that the $\Delta_3(L)$ should
be closer to the GUE result than to the GOE (which would be the appropriate
since the system is time reversible); this is exactly what is observed in a
quantum mechanical calculation \cite{Mart}.

A lot of the recent work in the area of quantum chaology
has been occupied with the
problem of relating {\em individual} eigenstates to the periodic orbits.
Due to the exponential proliferation of cycles this can not be pursued very
high up in the spectrum.  This problem is in itself very interesting.
But, if we
want to use semiclassical methods to investigate the transition from
classical to quantum mechanics, that is, if we are interested in
(asymptotically) high states, then we are bound to understand the
asymptotics of the set of periodic orbits.
The problem of exploring the asymptotics of this set is well examplified in
ref. \cite{Argam}.
It is the author's hope that this
paper provides some new ideas in this connection. \vspace{1cm}

Finally I would like to thank Predrag Cvitanovi\'{c}, Hans Henrik Rugh and
Viviane Baladi
for valuable
discussions. This work was supported by the Swedish Natural Science
Research Council (NFR).

\newcommand{\PR}[1]{{Phys.\ Rep.}\/ {\bf #1}}
\newcommand{\PRL}[1]{{Phys.\ Rev.\ Lett.}\/ {\bf #1}}
\newcommand{\PRA}[1]{{Phys.\ Rev.\ A}\/ {\bf #1}}
\newcommand{\PRD}[1]{{Phys.\ Rev.\ D}\/ {\bf #1}}
\newcommand{\PRE}[1]{{Phys.\ Rev.\ E}\/ {\bf #1}}
\newcommand{\JPA}[1]{{J.\ Phys.\ A}\/ {\bf #1}}
\newcommand{\JPB}[1]{{J.\ Phys.\ B}\/ {\bf #1}}
\newcommand{\JCP}[1]{{J.\ Chem.\ Phys.}\/ {\bf #1}}
\newcommand{\JPC}[1]{{J.\ Phys.\ Chem.}\/ {\bf #1}}
\newcommand{\JMP}[1]{{J.\ Math.\ Phys.}\/ {\bf #1}}
\newcommand{\JSP}[1]{{J.\ Stat..\ Phys.}\/ {\bf #1}}
\newcommand{\AP}[1]{{Ann.\ Phys.}\/ {\bf #1}}
\newcommand{\PLB}[1]{{Phys.\ Lett.\ B}\/ {\bf #1}}
\newcommand{\PLA}[1]{{Phys.\ Lett.\ A}\/ {\bf #1}}
\newcommand{\PD}[1]{{Physica D}\/ {\bf #1}}
\newcommand{\NPB}[1]{{Nucl.\ Phys.\ B}\/ {\bf #1}}
\newcommand{\INCB}[1]{{Il Nuov.\ Cim.\ B}\/ {\bf #1}}
\newcommand{\JETP}[1]{{Sov.\ Phys.\ JETP}\/ {\bf #1}}
\newcommand{\JETPL}[1]{{JETP Lett.\ }\/ {\bf #1}}
\newcommand{\RMS}[1]{{Russ.\ Math.\ Surv.}\/ {\bf #1}}
\newcommand{\USSR}[1]{{Math.\ USSR.\ Sb.}\/ {\bf #1}}
\newcommand{\PST}[1]{{Phys.\ Scripta T}\/ {\bf #1}}
\newcommand{\CM}[1]{{Cont.\ Math.}\/ {\bf #1}}
\newcommand{\JMPA}[1]{{J.\ Math.\ Pure Appl.}\/ {\bf #1}}
\newcommand{\CMP}[1]{{Comm.\ Math.\ Phys.}\/ {\bf #1}}
\newcommand{\PRS}[1]{{Proc.\ R.\ Soc. Lond.\ A}\/ {\bf #1}}

\newpage

\section*{Appendix}

We now demonstrate how to obtain a reccurence relation for the coefficients
of the expansion of
\begin{equation}
\prod_{k=0}^{\infty}(1+\frac{t}{\Lambda^k})
^{k+1}=
\sum_{n=0}^{\infty}b_n(\Lambda) t^n  \ \ .  \label{eqn:eulerk+1}
\end{equation}
Consider first the simpler problem of expanding the Euler product
\begin{equation}
\prod_{k=0}^{\infty}(1+\frac{t}{\Lambda^k})=
\sum_{n=0}^{\infty}a_n(\Lambda) t^n  \ \ !
\end{equation}
We extract the first factor
\begin{equation}
\prod_{k=0}^{\infty}(1+\frac{t}{\Lambda^k})=
(1+t)\prod_{k=1}^{\infty}(1+\frac{t}{\Lambda^k})=
(1+t)\sum_{n=0}^{\infty}a_n(\Lambda)(t/\Lambda)^n
\end{equation}
and obtain the recurrence formula
\begin{equation}
a_n=\frac{\Lambda^{1-n}}{1-\Lambda^{-n}}a_{n-1} \; \;
 \; \;  \; \; a_0=1  \ \ ,
\end{equation}
which can be exactly solved
\begin{equation}
a_n=\frac{\Lambda^{-\frac{n(n-1)}{2}}}
{\prod_{k=1}^{n}(1-\Lambda^{-k})}  \ \ .
\end{equation}
We now use the some technique for eq. \EqRef{eulerk+1}
\begin{equation}
\prod_{k=0}^{\infty}(1+\frac{t}{\Lambda^k})^{k+1}=
\prod_{k=0}^{\infty}(1+\frac{t}{\Lambda^k})\cdot
\prod_{k=1}^{\infty}(1+\frac{t}{\Lambda^k})^{k}=
\left( \sum_{n=1}^{\infty} a_n(\Lambda) t^n \right)
\left( \sum_{n=1}^{\infty} b_n(\Lambda) (t/\Lambda)^n \right) \ \ ,
\end{equation}
leading to the reccurence relation
\begin{equation}
b_n=\frac{1}{1-\Lambda^{-n}}\sum_{l=0}^{n-1}a_{n-l}b_{l}\Lambda^{-l}
 \; \;  \; \;  \; \; b_0=1 \ \ .
\end{equation}

\newpage
\section*{Figure captions}

\FigCap{1}{The spectrum (a) of model A ($\diamond$)
and B (+) and its asymptotic
expression ($\Box$)
and (b) the trace for model A (full line) and
B (dashed). The asymptotic expression has been used to calculate its
discrete contribution $\Psi_{zeros}$ to the trace.}

\FigCap{2}{When $Z(k)$ has a branch cut along the positive imaginary axis
the Fourier transform of the logaritmic derivative has to be integrated
along $\gamma_{branch}$.}

\FigCap{3}{The trace for model C}

\FigCap{4}{$p(\Delta)$ (a), the spectrum (b), and the trace (c) for model D
with different perturbation strengths $\epsilon$.}

\FigCap{5}{Members of some simple sequences in the hyperbola billiard}

\FigCap{6}{Members of some composed sequences in the hyperbola billiard}

\FigCap{7}{Graphical representation of the allowed cycles in some composed
sequences}

\FigCap{8}{Spectrum of the hyperbola billiard for various numbers of
pseudo orbits $N_{cut}$.}

\FigCap{9}{Location of the zero $k_0$ as a function of $N_{cut}$.}

\FigCap{10}{The trace for the hyperbola billiard using direct summation
(eq. 2) (long dashed),
and as a sum over eigenvalues, with $N_{cut}=200$ (short dashed),
and $N_{cut}=460$ (full line).}

\FigCap{11}{$p(\Delta)$ in the hyperbola billiard obtained by numerical
simulation}

\newpage
\section*{Tables}

\begin{table}[h]

  \centering
  \begin{tabular}{|l|c|c|c|} \hline
 & $j_{max}$ & $T(j_{max})$ & $\Lambda (j_{max})$  \\
\hline
$210^j$ & 3 & 7.13$\equiv T_a$ & 19.65 \\
$21210^j$ &16 & 18.45$\equiv T_b$ & 527.9  \\
$220^j$ & 51 & 27.61$\equiv T_c$ & 922.9  \\
$2210^j$ & $\sim$640 & $\sim$92$\equiv T_d$ & $\sim$34000 \\
    \hline
$20^j$ & $\infty$ &- &-   \\
$110^j$ & $\infty$ & - & -  \\
\hline
  \end{tabular}

\caption{Some simple sequences of cycles. Invariants are given
for last
cycle in each sequence} \end{table}

\end{document}